# Diameter Dependence of the Transport Properties of Antimony Telluride Nanowires

Yuri M. Zuev[a], Jin Seok Lee[b†], Clément Galloy[c], Hongkun Park[b], and Philip Kim[a,c,*]

[a]Department of Applied Physics and Applied Mathematics, Columbia University, New York, NY 10027, USA; [b]Department of Chemisty and Chemical Biology and Department of Physics, Harvard University, Cambridge, MA 02138, USA; [c]Department of Physics, Columbia University, New York, NY 10027, USA

[†]Present Address: Department of Chemistry, Sookmyung Women's University, Seoul 140-742, Korea

*To whom correspondence should be addressed: pk2015@columbia.edu

**Abstract**

We report measurements of electronic, thermoelectric, and galvanomagnetic properties of individual single crystal antimony telluride ($Sb_2Te_3$) nanowires with diameters in the range of 20-100 nm. Temperature dependent resistivity and thermoelectric power (TEP) measurements indicate hole dominant diffusive thermoelectric generation, with an enhancement of the TEP for smaller diameter wires up to 110 µV/K at $T$ = 300 K. We measure the magnetoresistance, in magnetic fields both parallel and perpendicular to the nanowire [110] axis, where strong anisotropic positive magnetoresistance behavior was observed.

**Keywords** : Thermoelectric power, $Sb_2Te_3$, topological insulator

Maximizing the thermoelectric figure of merit, ZT, is of critical importance for any potential commercial usage of thermoelectric energy conversion. The best bulk thermoelectric materials have a ZT of 1, which has not changed for the last half century[1]. Nanostructured materials have shown a possibility to achieve a ZT > 1, both experimentally[2,3,4,5,6] and theoretically[7,8]. It is predicted that quantum confinement in Bi nanowires (NWs) would lead to a semimetal-semiconductor transition as well as large gains in ZT for wires below ~50nm, depending on the crystallographic orientaiton[7,9]. Several studies have achieved a $ZT = \frac{S^2 \sigma T}{\kappa} > 1$ in Si nanowires (here S (Seebeck coefficient) is the thermoelectric power (TEP), $\sigma$ is the electrical conductivity, $\kappa$ is the thermal conductivity and T is the temperature) by increasing phonon boundary scattering and therefore decreasing the thermal conductivity term[3,10]. Although minimizing the thermal conductivity is one strategy to maximize ZT, maximizing the electronic contribution, i.e., the power factor, $S^2\sigma$, is another[8,11].

In this letter, we investigate how nanowire diameter influences the electronic contribution to the thermoelectric figure of merit in single crystal chalcogenide $Sb_2Te_3$ NWs. Antimony telluride ($Sb_2Te_3$) is a small bandgap semiconductor with a gap of 0.28 eV[11,12] (recent calculations[13,14] show gap ≈ 0.1 eV). Due to antistructural defects in the rhombohedral crystal structure, $Sb_2Te_3$ is generally *p*-type with the Fermi level deep in the double valence band. To obtain a better understanding of the electronic structure, we investigate the temperature and carrier density dependence of the resistivity and TEP as a function of NW diameter. Antimony telluride is also a topological insulator with a bulk energy gap and topologically protected surface states[13,14]. It has been reported[15] that the application of a magnetic field parallel to the topologically insulating nanowires ($Bi_2Se_3$ NWs) produces Aharonov-Bohm oscillations in the magnetoresistance which stems from the surface states. Although the surface states of $Sb_2Te_3$ are more difficult to access than those in $Bi_2Se_3$ due to the small energy gap, we nevertheless investigate both parallel and perpendicular magnetoresistance effects, thereby testing the possibility of observing galvanomagnetic quantum interference effect in $Sb_2Te_3$.

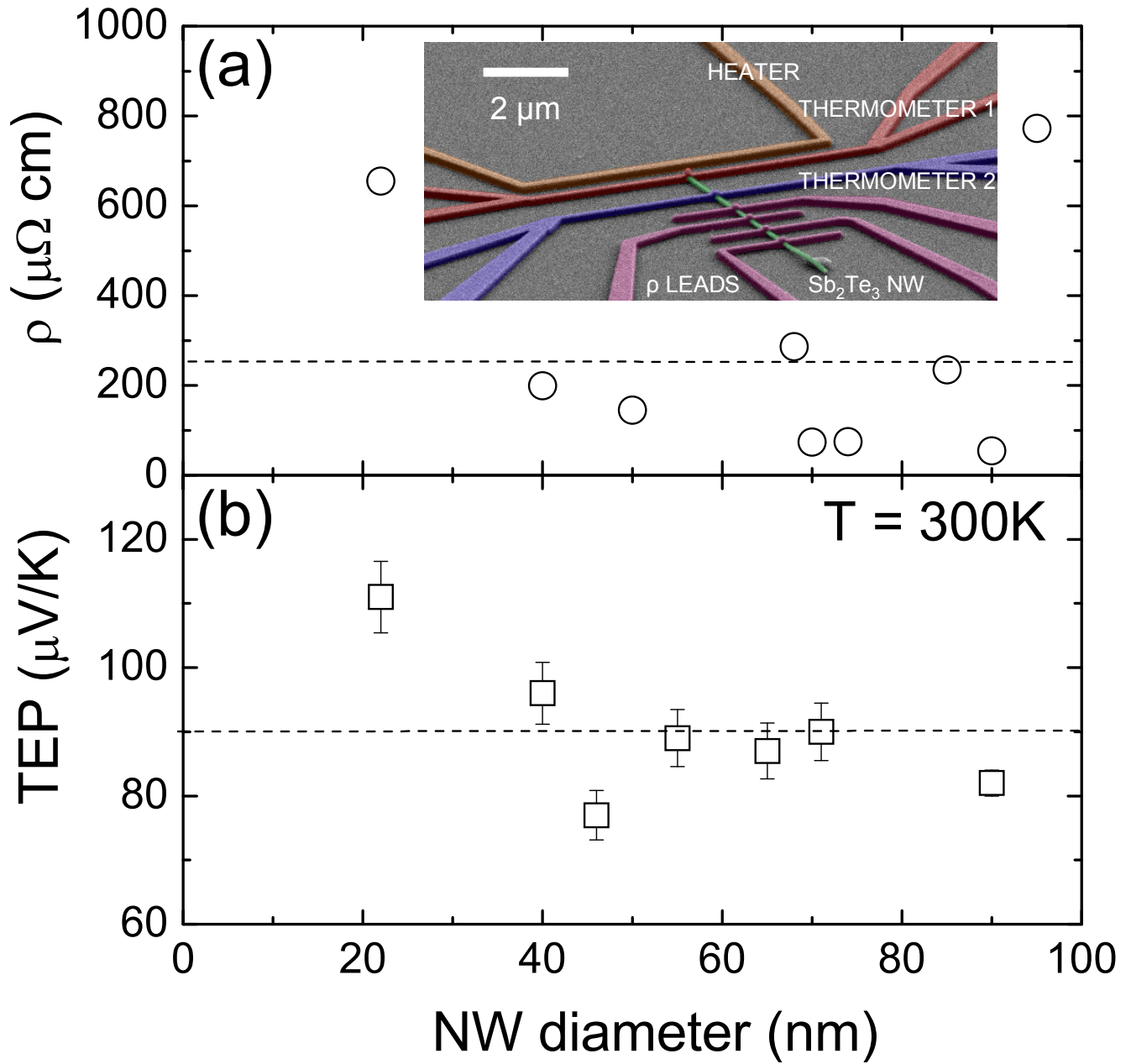

**Figure 1** (a) Resistivity and (b) thermopower as a function of NW diameter at $T = 300$K. Error bars in (a) are smaller than data markers. Inset shows a false color SEM image of a typical device. Thermopower is enhanced in smaller wires. The dashed lines in (a) and (b) are bulk values from References 12,18, and 19.

Single crystal $Sb_2Te_3$ nanowire samples were synthesized using the vapor-liquid-solid method described in Reference 16. The crystallographic orientation of the nanowire axis was measured to be in the [110] direction[16]. Details of nanowire characterization, including atomic resolution images and chemical elemental mapping, are shown in the Supporting Information section Figure S1. Substrates containing nanowires were mildly sonicated in isopropyl alcohol solution and deposited onto 950 nm $SiO_2$/Si wafers. A histogram of the diameter distribution is shown in Figure S2. The carrier density inside the nanowires was adjusted by applying a gate voltage ($V_g$) to the degenerately doped Si substrate. Ohmic contact to the NWs was made with standard electron beam lithography followed by evaporation of Ti/Ni (2/200nm) and lift-off procedure. No etching step was required prior to metal evaporation in order to make Ohmic contact. Atomic Force Microscopy (AFM) was used to measure the NW diameter with accuracy to within 1 nm.

The TEP measurement technique was described in detail elsewhere[17]. In brief, a controlled temperature gradient $\Delta T$ was applied to the sample by a microfabricated heater while the resulting thermally induced voltage $\Delta V$ was measured by the voltage probes to acquire the TEP, $S = -\Delta V/\Delta T$ (see the upper inset of Fig. 1 for device layout). Since the $SiO_2$ substrate is insulating the measured $\Delta V$ reflects the intrinsic properties of the nanowire. Local temperature variations were measured with two metal four-probe microthermometers. The condition of linear response regime, $\Delta T \ll T$, was always satisfied. The resistivity of the nanowire was measured using a four-terminal geometry in order to get rid of contact resistance. The resistivity and TEP measurements were done in the DC configuration, while the magnetoresistance measurements were done with a lock-in amplifier at low frequency.

Figure 1 shows the electronic and thermoelectric properties of the nanowires measured at room temperature, T = 300 K, as a function of NW diameter. Bulk values[12,18,19] for $Sb_2Te_3$ resistivity = 250 $\mu\Omega \cdot cm$ and TEP = 90 $\mu V/K$ are plotted as dashed lines in Fig. 1a and 1b, respectively. The typical resistivity for $Sb_2Te_3$ NWs is comparable to the bulk value and does not exhibit strong diameter dependence. As shown in Figure 1b, however, the thermoelectric power is enhanced in smaller diameter

NWs: TEP of a 22-nm NW is measured to be 111 μV/K while that of a 95-nm NW is 81 μV/K, indicating more than 30 % TEP enhancement in the smallest NWs we measured.

The temperature dependence of the resistivity and TEP exhibits clear changes in the transport behavior as a function of NW diameter. Figure 2a shows the normalized four-terminal resistance plotted as a function of temperature on a semilog plot from T = 300 K to T = 5 K. Error bars are taken as the standard deviation of consecutive measurements and are smaller than the data markers where they are not seen. Between 300 K and ~50 K the resistance decreases linearly as a function of decreasing temperature due to a decreasing phonon population, similar to bulk metals[20]. Below ~50 K, NWs with diameters larger than 45 nm exhibit resistance saturation because the scattering is dominated by boundary and impurity scattering. For NWs smaller than 45 nm the resistance increases to several times the room temperature value. We note that there are some sample to sample variations due to mesoscopic fluctuations of the defect arrangement. The exact nature of this low temperature non-metallic behavior of small diameter NWs is not clear presently. We note, however, that $Sb_2Te_3$ can undergo reversible phase change between an amorphous and a crystalline phase by the application of a voltage pulse[16]. For small diameter NWs it was often necessary to anneal the devices by flowing large current in order to improve the overall conductance (Fig. 2a inset). One potential scenario for resistance increase is that some small number of amorphous grain boundaries remain in small diameter NWs and contribute to resistance increase at low temperature. If one grain became highly resistive at low temperatures, the overall resistance of the nanowire will become high.

We now turn our attention to the temperature dependence of TEP. For all NWs we measured, the profile of the TEP is linear to sublinear as a function of temperature (Fig. 2b). The thermoelectric coefficient is proportional to the asymmetry of the density of states, and the sign of the TEP is proportional to the majority charge carrier[20]. The semiclassical Mott relation for diffusive thermopower is given by $S = -\dfrac{\pi^2 k_B^2 T}{3e} \cdot \dfrac{d \ln \sigma(E)}{dE}\bigg|_{E_F}$, and predicts a linear temperature dependence for the TEP[20].

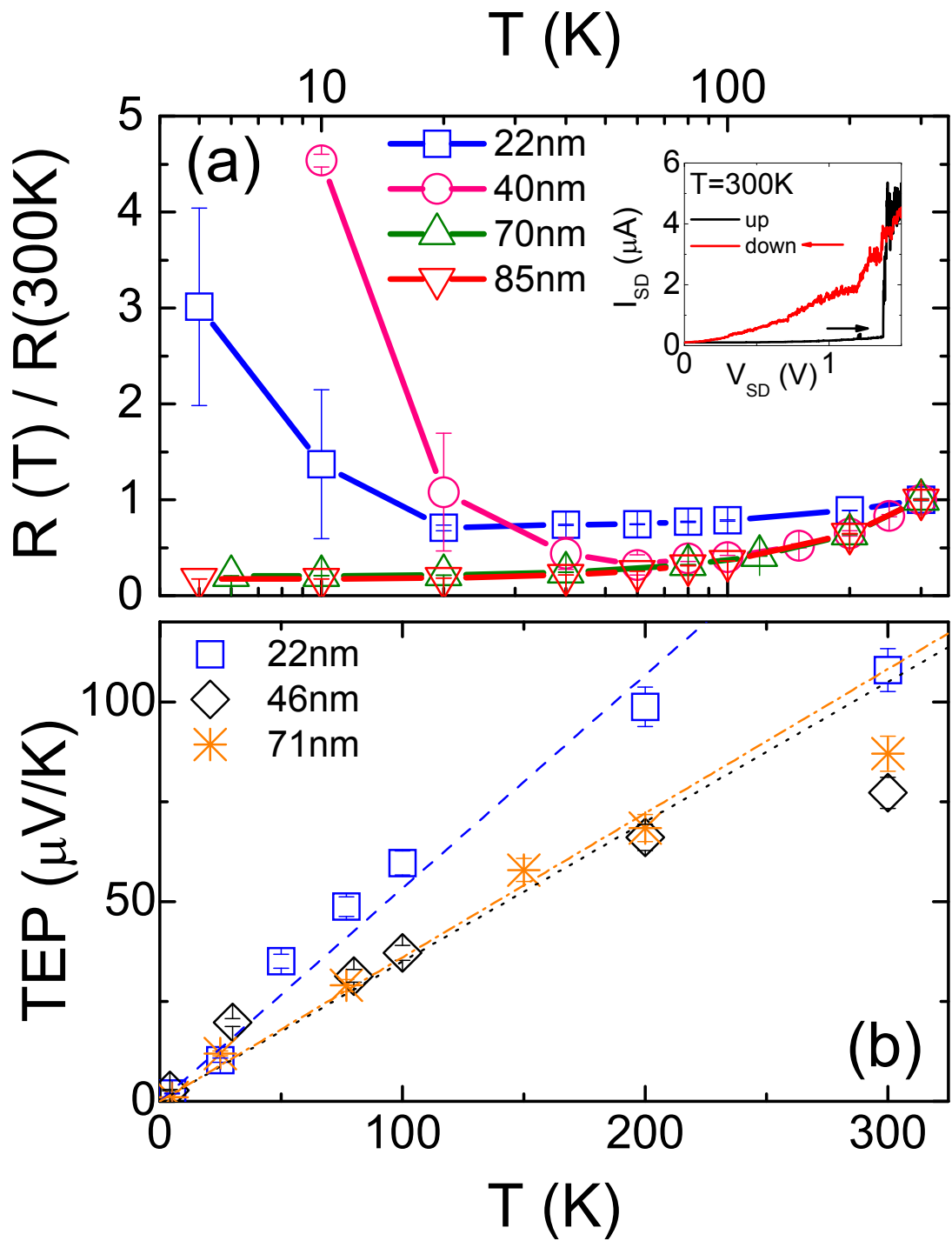

**Figure 2** (a) Normalized 4-terminal resistance and (b) thermopower as a function of temperature for several NWs. The same 22nm diameter NW is seen in (a) as in (b). The inset in (a) shows the current versus bias voltage characteristics before and after crystallization of the 22 nm NW by application of a high bias. The linear fits in (b) correspond to diffusive thermopower generation, and are discussed in the text.

Here $e$ is the charge of an electron, $k_B$ is the Boltzmann constant, $T$ is the temperature, $\sigma(E)$ is the energy dependent conductivity, and $E_F$ is the Fermi energy. Consequently, the positive linear temperature dependence below $T = 200$ K in $Sb_2Te_3$ NWs signifies hole dominant diffusive thermoelectric generation. Assuming a simple parabolic band structure, the slope of the linear fit can be used to extract the carrier density in the nanowires[21,22,23]. Provided that bulk hole effective mass[12] of $m^* = 0.78m_e$ ($m_e$ is the electron rest mass) can still be used for holes in nanowires, we estimate a hole carrier density on the order of $10^{20}$ cm$^{-3}$ in our 70-nm nanowires and the corresponding mobility of ~800 cm$^2$/Vs. This carrier density values are comparable to values published for bulk $Sb_2Te_3$[12,24,25]. The carrier density decreases linearly as a function of decreasing diameter, and is smaller by a factor of 2 or so in the 22 nm diameter NWs. It is possible that the decreasing TEP is the result of the decreasing density in the smaller diameter nanowires. It is noted that the measured TEP values at T = 300K are identical before and after the recrystallization process, even thought the resistance changes by two orders of magnitude. In a bulk sample a Hall measurement can be used to determine the carrier density. In nanowires, however, such a measurement cannot be performed due to their 1-dimensionality. Since the electron density predicted with the electric field effect can often overestimate the actual carrier density, the technique described above provides a useful experimental tool[21].

The application of $V_g$ enables the control over the carrier density, and thus the position of the Fermi energy, and can be used to achieve high ZT values[9,10,23]. In spite of the high hole carrier density (~$10^{19}$ cm$^{-3}$), the smallest diameter NWs ($d = 22$ nm) show $V_g$-dependent transport behavior. Using the standard cylinder on plane model, we estimate that the induced field effect density in the 22 nm NW for the applied 100 V of gate voltage is ~4 x $10^{19}$/cm$^3$. In Fig. 3a, we plot the two-terminal current ($V_{SD} = 50$ mV) as a function of $V_g$ and temperature. At room temperature, the source-drain current $I_{SD}$ decreases ~ 50 % as $V_g$ changes from -40 V to 60 V, confirming that the majority carriers are holes. As $T$ decreases, this $p$-type behavior becomes more pronounced, and the conduction turns off for $V_g > 20$ V at $T < 50$ K.

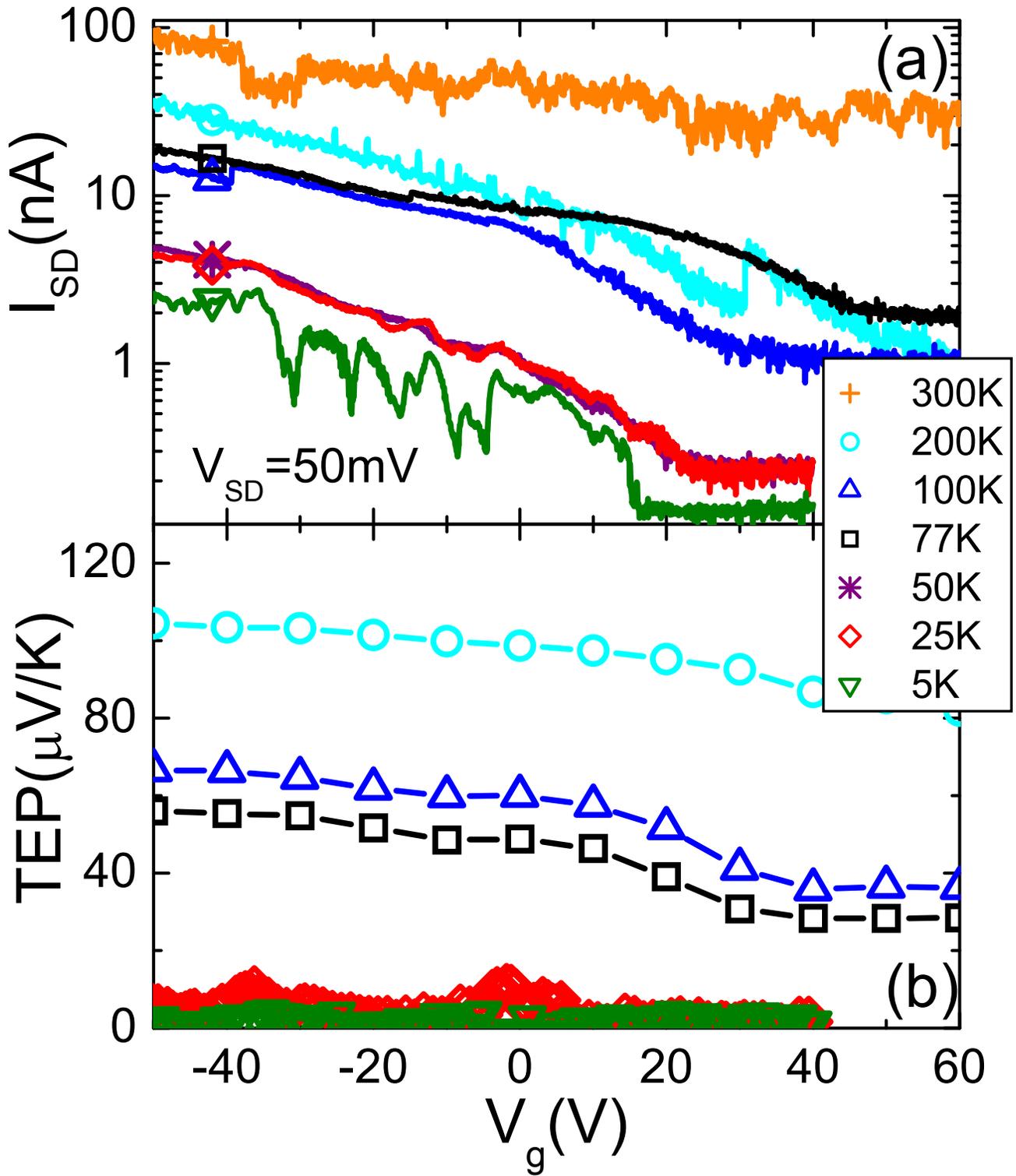

**Figure 3** (a) Gate dependence of the source-drain current ($V_{SD}$ = 50 mV) and (b) thermopower of a 22 nm diameter NW at different temperatures. The NW exhibits quantum oscillation at low temperatures, T < 30K. It also shows p-type behavior, turning off for positive gate voltages. The thermopower varies only slightly as a function of gate voltage.

At the low temperature of $T \sim 5$ K, the conductance shows mesoscopic fluctuations due to a disorder driven Coulomb blockade or universal conductance fluctuations. This observation suggests that the phase coherence length in nanowires is on the order of the channel length $\sim 1.5$ μm at this temperature.

The applied $V_g$ modulates not only the nanowire conductance but also TEP. Fig. 3b shows that the gate modulated TEP decreases monotonically as $V_g$ increases. At lower temperatures (T < 100 K) where the device conductance exhibits significant turn-off behavior (on-off ratio > 10), the TEP exhibits a developing step in $V_g$. It is interesting to note that even in the turn-off regime of the device, the TEP shows a finite positive value, indicating holes are responsible for the remnant charge transport in this regime. The TEP in all $V_g$ regimes in our experiment decreases with decreasing $T$, suggesting the carriers responsible for the observed TEP are still generated rather than activated[26].

Finally, we discuss the magnetoresistance measurements performed on the nanowires. Antimony telluride has a non-spherical Fermi surface consisting of 6 ellipsoids tilted at an angle to the basal plane, where two valence bands, upper and lower, are responsible for conduction[12,24]. The dependence of the perpendicular magnetoresistance up to B = ±10 T of a 70-nm NW is parabolic at all temperatures (Figure 4a). In a two band model the magnetoresistance is given by $\rho(B) = \rho_{B=0}(1 + A \cdot B^2)$, where A is a constant. In the inset to Figure 4 we fit the data from B = -5 to 5 T with a second order polynomial, extracting the parameter $A$ that decreases with increasing temperature. Similar magnetoresistance dependence has been observed in Bi NW[27] and bulk $Sb_2Te_3$[24,28]. The magnetoresistance is slightly subparabolic for $B > 5$ T. For larger ($d > 50$nm) nanowires, the carrier mobility decreases (as signified by the resistance increase) as the temperature is raised because of increased phonon scattering. As the mobility decreases, the mean free path also decreases and the magnetoresistance changes less. We note that the nanowire samples do not exhibit Shubnikov-de Hass oscillations, most likely due to low mobility. Signatures of topological surface states are not observed because the Fermi level is deep in the bulk valence band away from the gap, even at the highest $V_g$. Figure 4b shows a comparison between the magnetoresistance for a 70-nm NW at 6 K when the magnetic field is in the direction perpendicular and

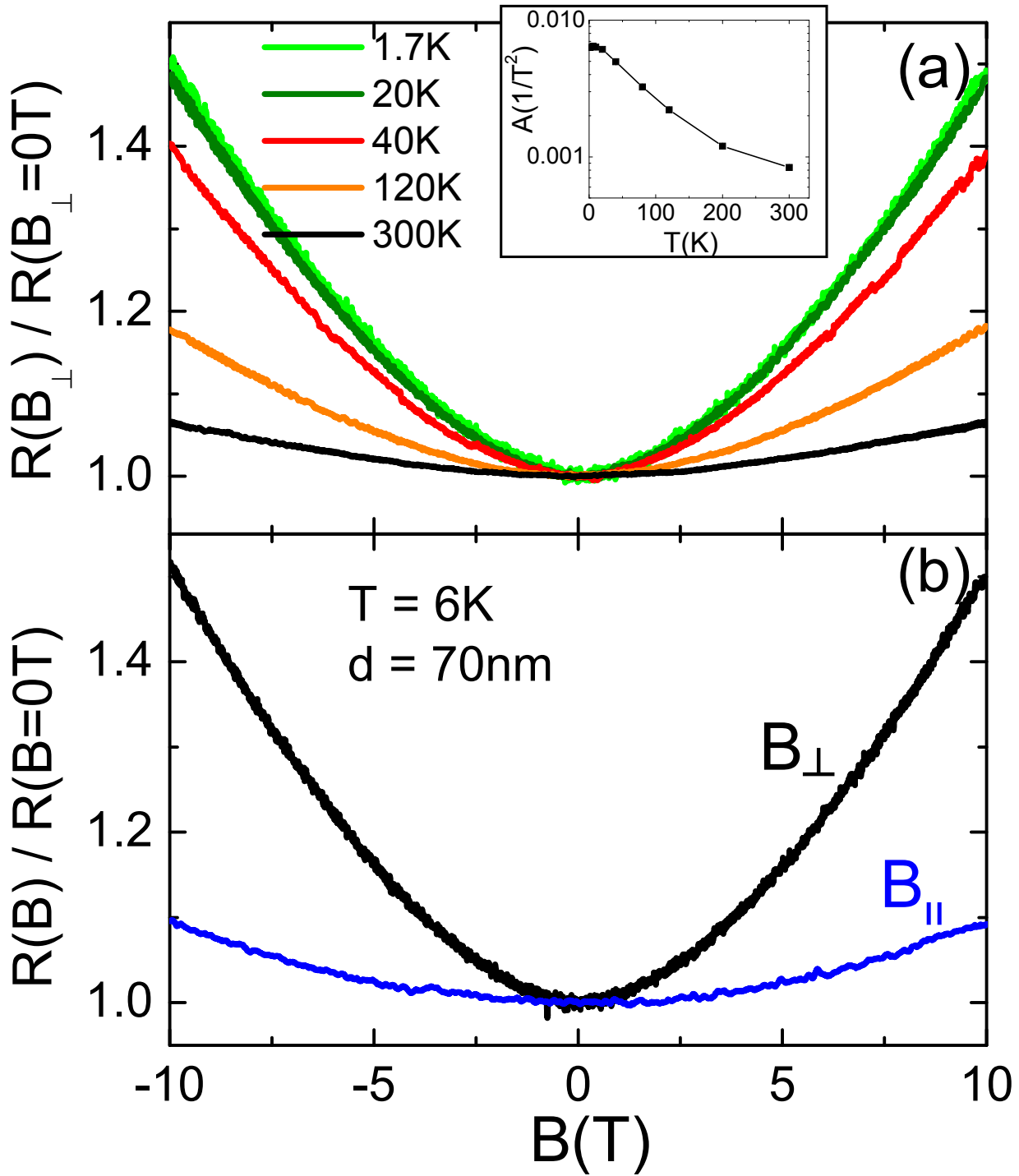

**Figure 4** (a) Temperature dependence of normalized magnetoresistance in a perpendicular magnetic field of a 70nm diameter NW. The parabolic behavior is attributed to the contribution of holes from two valence bands to the magnetoresistance. The inset shows the decrease of the curvature as the temperature increases. (b) Comparison of the magnetoresistance for a magnetic field applied parallel and perpendicular to the NW [110] axis at *T* = 6K.

parallel to the NW [110] axis. While the perpendicular magnetoresistance increases by 50 % at B = 10 T the parallel magnetoresistance increases by only 10 %, signifying the anisotropy of the Fermi surface in $Sb_2Te_3$. If the electron orbits are closed, the magnetoresistance should saturate at high fields[20]. The absence of saturation in our nanowires in both parallel and perpendicular fields suggests that electrons are traversing open orbits along the Fermi surface. The effective mass of the upper and lower valence bands in $Sb_2Te_3$ varies from $m^* = .034m_e$ to $m^* = 1.24m_e$ due to the anisotropy of the ellipsoidal hole pockets[12]. This anisotropy of the effective mass tensor might be responsible for the magnetoresistance variations in parallel and perpendicular fields. More detailed studies of the cyclotron resonances along different directions of the crystal axes is needed to provide a more detailed understanding of the nanowire Fermi surface.

In conclusion, we have measured the resistivity and thermoelectric power of antimony telluride nanowires as a function of temperature, nanowire diameter, and gate voltage. The resistivity values are close to the bulk value (250 $\mu\Omega$·cm) while the TEP shows enhancement in smaller diameter nanowires by up to ~30%. It is found that the thermoelectric power factor is not greatly enhanced in smaller diameter nanowires in the measured diameter range. For larger diameter NWs, the resistivity shows metallic linear behavior above T ≈ 50 K and exhibits saturation below T ≈ 50 K. Both the gate voltage dependence and TEP measurements indicate hole dominant diffusive thermoelectric generation. The magnetoresistance is parabolic in magnetic field and does not saturate even at high perpendicular and parallel fields.

**Acknowledgements.** This work was financially supported by the MRSEC Program of the NSF Grant No. DMR-02-13574, and NSF CAREER Grant No. DMR-0349232.

# Supplementary information

## 1. Atomic Structural and Chemical Compositional Characterization

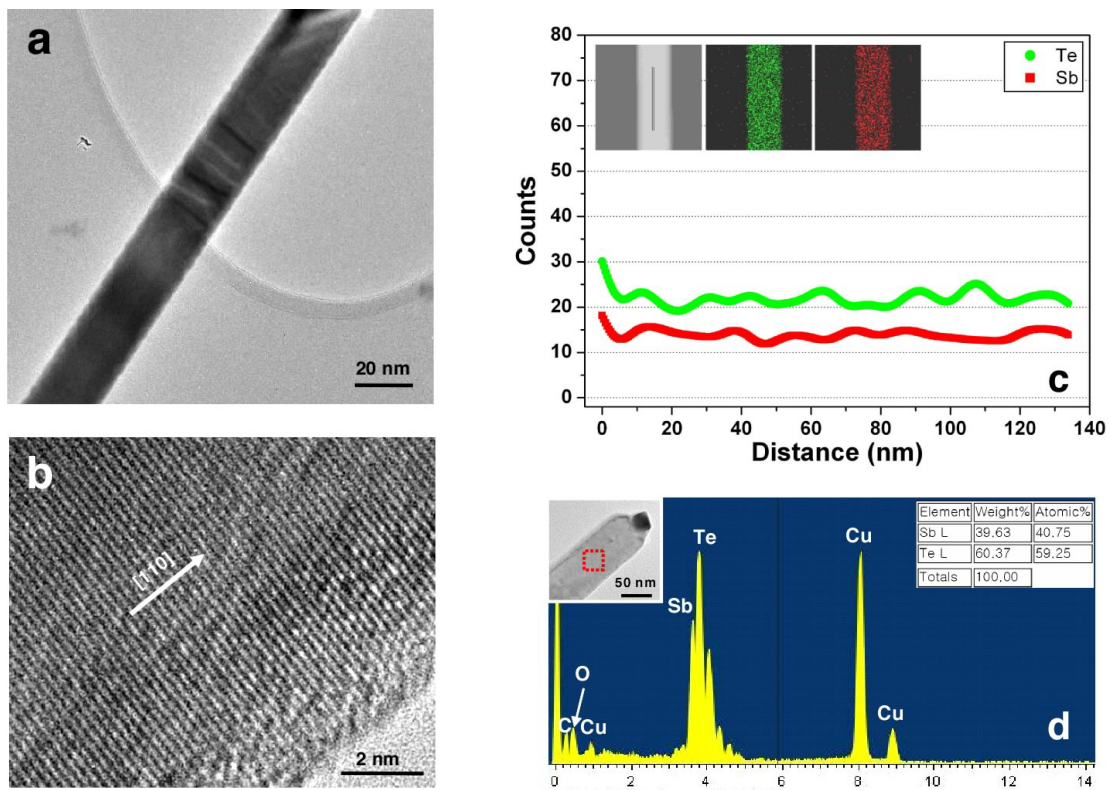

**Figure S1.** Hig resolution transmission electron microscope (HRTEM) images of a typical $Sb_2Te_3$ nanowire are shown in (a) and (b). (c) Cross sectional Energy-dispersive X-ray spectroscopy (EDX) elemental line scans of a $Sb_2Te_3$ nanowire taken along the axis. The inset shows bright field scanning-mode TEM images with the corresponding EDX elemental mapping. (d) EDX spectrum showing that the atomic ratio of Sb to Te is constant throughout the wire.

## 2. Nanowire Diameter Distribution

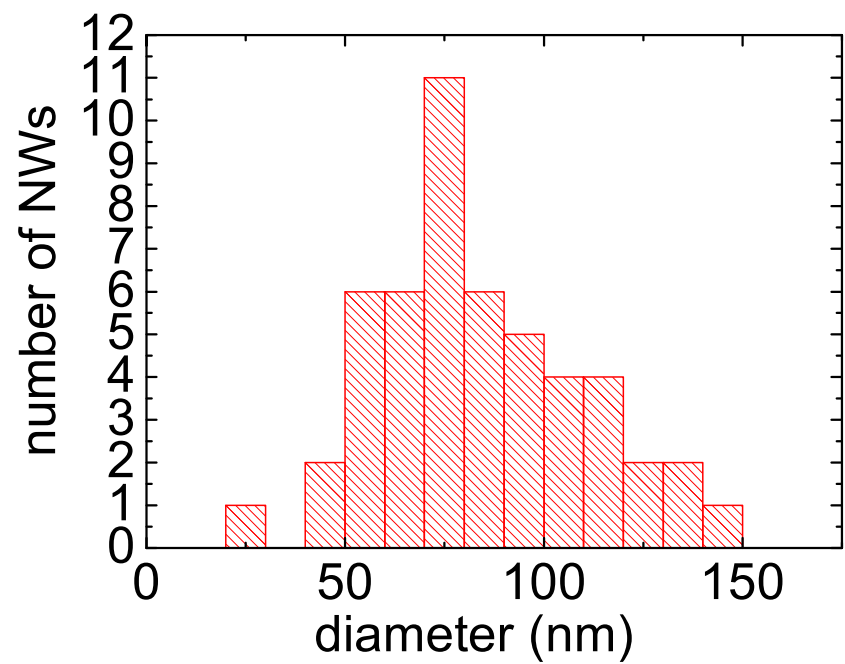

**Figure S2.** A histogram of the diameter (measured with AFM) distribution of nanowires located after the NWs were dispersed in solution.